\documentclass[twocolumn]{aastex631}
\usepackage{amsmath}
\usepackage{graphicx}
\usepackage{hyperref}
\usepackage{booktabs}

\defcitealias{Cehula+24}{C24}

\newcommand{\be}{\begin{equation}}
\newcommand{\ee}{\end{equation}}

\shorttitle{Suppression of shock X-ray emission in novae from turbulent mixing with cool gas}
\shortauthors{B.~D.~Metzger et al.}

\begin{document}

\title{Suppression of shock X-ray emission in novae from turbulent mixing with cool gas}

\author[0000-0002-4670-7509]{Brian D.~Metzger}
\affil{Department of Physics and Columbia Astrophysics Laboratory, Columbia University, New York, NY 10027, USA}
\affil{Center for Computational Astrophysics, Flatiron Institute, 162 5th Ave, New York, NY 10010, USA}

\author[0000-0002-0041-4356]{Lachlan Lancaster}
\affil{Department of Astronomy, Columbia University, 550 W 120th St, New York, NY 10025, USA}
\affil{Center for Computational Astrophysics, Flatiron Institute, 162 5th Ave, New York, NY 10010, USA}\thanks{Simons Fellow}

\author[0000-0002-6679-0012]{Rebecca Diesing}
\affil{School of Natural Sciences, Institute for Advanced Study, Princeton, NJ 08540, USA}
\affil{Department of Physics and Columbia Astrophysics Laboratory, Columbia University, New York, NY 10027, USA}

\correspondingauthor{Brian Metzger}
\email{bdm2129@columbia.edu}

\begin{abstract} 

Shock interaction in classical novae occurs when a fast outflow from the white dwarf $\gtrsim 1000$ km s$^{-1}$ collides with a slower, cooler shell of gas released earlier in the outburst. The shocks radiate across the electromagnetic spectrum, from radio synchrotron to GeV gamma-rays. The hot shocked gas also emits $\gtrsim $ keV thermal X-rays, typically peaking weeks after the eruption, once the ejecta becomes transparent to photoelectric absorption. However, the observed hard X-ray luminosities are typically $\gtrsim 4$ orders of magnitude smaller than would be naively expected given the powerful shocks implied by the gamma-rays. We argue that a key missing piece to this puzzle is turbulence behind the shock, driven, e.g., by thin-shell and/or thermal instabilities. Turbulence efficiently mixes the hot X-ray emitting gas with cooler gas, sapping the hot gas of energy faster than it can directly radiate. Using analytic arguments motivated by numerical simulations, we show that energy losses due to turbulent mixing can easily balance shock heating, greatly reducing the volume of the hot gas and suppressing the X-ray luminosity.  Equating the characteristic thickness of the X-ray emitting region to the minimum outer length scale of the turbulence capable of cooling the hot gas through mixing, we obtain X-ray luminosities consistent with nova observations if only $\sim 1\%$ of the shock's kinetic power goes into turbulent motions. A similar process may act to suppress thermal X-rays from other shock powered transients, such as interacting supernovae.

\end{abstract}


\section{Introduction}

Novae are sudden outbursts driven by runaway nuclear burning on the surfaces of white dwarfs of hydrogen-rich material accreted from a binary companion \citep{Gallagher&Starrfield76}.  They reach peak visual luminosities $\sim 10^{4}-10^{5}L_{\odot}$ and eject large quantities of mass $\sim 10^{-6}-10^{-4}M_{\odot}$ at high velocities $\gtrsim 10^{3}$ km s$^{-1}$. Though conventionally believed to be powered directly by nuclear energy, growing evidence suggests that shocks also play a decisive role in powering nova emission across the electromagnetic spectrum \citep{Chomiuk+20}.  This evidence includes multiple velocity components in their optical spectra exhibiting complex time evolution (e.g., \citealt{Friedjung87,Williams&Mason10,Aydi+20b}); thermal X-ray emission weeks after the outburst which is too hard to arise from the burning layer on the white dwarf surface (e.g.,~\citealt{Mukai+08,Orio+21,Gordon+21,Sokolovsky+22}); and early maxima in their radio light curves which far exceed that from $\sim 10^{4}$ K photo-ionized ejecta (e.g.,~\citealt{Chomiuk+14,Weston+16a,Finzell+18,Chomiuk+21}).  

The clearest indicator of powerful shocks in novae came with the discovery by {\it Fermi} LAT of $\gtrsim$ 100 MeV gamma-rays, observed coincident within a few days of the optical peak and lasting weeks \citep{Ackermann+14}. The first gamma-ray nova occurred in the symbiotic binary V407 Cyg \citep{Abdo+10}, suggesting shocks produced by the nova outflow colliding with the dense wind of the companion giant \citep{Martin&Dubus13}.  However, in the ensuing decade and a half, gamma-rays have been detected from over 20 classical novae with main sequence or moderately evolved companions \citep{Ackermann+14,Cheung+16,Franckowiak+18}.  Remarkably, the outflows in classical novae must be colliding with dense gas immediately surrounding the binary, inconsistent with the observed environments of cataclysmic variables (e.g., \citealt{Hoard+14}). The required dense gas instead likely represents slower matter ejected earlier in the outburst, thus implicating the shocks as being {\it internal} \citep{Lloyd+92,Metzger+14,Martin+18,Hachisu&Kato22}.  The sudden appearance of high-velocity outflows may indicate the delayed onset of a fast wind from the white dwarf driven by radiation pressure on the iron opacity bump \citep{Kato&Hachisu94,Shen&Quataert22,Kato+22}.  This can occur once enough mass has been removed from the envelope, likely due in part to interaction with the orbiting binary companion (e.g., \citealt{Shankar+91,Chomiuk+14}).

The GeV gamma-rays in novae are mainly produced by collisions of cosmic ray protons accelerated at the shock with ambient protons \citep{Li+17,Vurm&Metzger18,Martin+18,Diesing+23}, while shock-accelerated electrons power the synchrotron radio emission (e.g., \citealt{Weston+16a,Vlasov+17}). Novae thus provides a new probe of relativistic particle acceleration in non-relativistic shocks, complementary to those in other astrophysical environments (e.g., \citealt{Metzger+15,Metzger+16,Diesing+23}).  

The high densities in nova outflows result in short radiative cooling times behind the shock, leading to complex multi-dimensional and multi-phase behavior \citep{Steinberg&Metzger18} and the formation of a cool and dense shell of gas \citep{Metzger+14} and, ultimately, dust \citep{Gehrz88,Derdzinski+17,Chong+25}. This thin shell is susceptible to an instability \citep{Vishniac94}, whereby lateral perturbations of the interaction front cause material to be diverted from convex to concave regions, in such a way that oppositely directed flows create elongated regions, imparting the cool gas shell and shock front with a ``corrugated'' geometrical structure \citep{Stevens+92,Walder&Folini00,Pittard09,Parkin+11,McLeod&Whitworth13,Kee+14}.  On top of any intrinsic temporal or directional variability in the white dwarf outflow, these instabilities drive turbulence in the post-shock gas \citep{Steinberg&Metzger18}.

The high densities in nova outflows also render the gas both upstream of the shock, and in the downstream cooling layer, largely neutral. These regions efficiently absorb UV and X-ray radiation from the shock, reprocessing its kinetic power to optical wavelengths, where the radiation escapes due to the lower opacity \citep{Metzger+14}. Comparing the gamma-ray and optical luminosities of novae leads one to deduce that an order-unity fraction of nova optical emission is shocked-powered \citep{Metzger+14,Chomiuk+20}, similar to ``interaction powered'' supernovae (e.g., \citealt{Smith+07}). This inference is supported by temporal correlations observed in a few well-studied novae, ASASSN-16ma and ASASSN-18fv, between the gamma-ray and optical light curves (e.g., \citealt{Li+17,Aydi+20}), as would be expected if both cosmic-ray acceleration and reprocessed thermal radiation track the instantaneous shock power, and hence each other \citep{Metzger+14}. However, classical novae exhibit a wide range of gamma-ray luminosities (e.g., \citealt{Franckowiak+18}), and the relative importance of shocks to the optical emission likely shows considerable diversity across the population. 

The high velocities of nova shocks $\gtrsim 1000$ km s$^{-1}$ (e.g., \citealt{Aydi+20b}) imply the shocked gas is heated to temperatures $T \gtrsim 10^{7}$ K.  Naively, one would therefore predict the emitted radiation should occur (prior to any absorption) as $\sim$keV X-rays carrying a luminosity comparable to the shock's power.  The latter can in turn be probed by the nova optical luminosity, if a significant fraction of the optical emission is indeed powered by reprocessed shock emission. It is usually not possible to directly test this expectation for several weeks after the outburst because the X-ray signal is strongly attenuated by photoelectric absorption. However, as the ejecta expands and the column of absorbing gas outside of the shock drops, eventually the $\gtrsim$ keV X-ray emission rises on a characteristic timescale of several weeks to months \citep{Mukai&Ishida01,Mukai+08,Vlasov+16,Gordon+21}. As a result of the rapidly decreasing opacity with photon energy, such detections become possible first at hard X-ray energies $\gtrsim$ 10 keV, such as probed by {\it NuSTAR} observations (e.g., \citealt{Nelson+19}).

But in contradiction to expectations, once revealed, the intrinsic (absorption corrected) hard X-ray luminosities of novae are measured to be only $\mathcal{L}_{\rm X} \sim 10^{33}-10^{34}$ erg s$^{-1}$ \citep{Nelson+19,Sokolovsky+20,Gordon+21,Sokolovsky+22}. These are typically $\sim 100$ times lower than the gamma-ray luminosities, and a factor $\gtrsim 10^{4}$ lower than expected if the shocks were to place $\lesssim 1\%$ of their kinetic power into gamma-rays (the ratio one infers from the correlated optical/gamma-ray emission; e.g., \citealt{Aydi+20}). The extremely weak shock X-ray emission from classical novae presents one of the biggest challenges to understanding these systems \citep{Chomiuk+20}.

Unexpectedly weak X-rays from high velocity shocks is not a phenomenon exclusive to novae.  X-ray deficits are also seen from the wind-blown bubbles of massive stars \citep{Weaver+77}, where the measured X-ray luminosities are a factor $\sim 100$ times lower than would be expected based on the injected power from the stellar winds (e.g., \citealt{Dunne+03,Townsley+03,Lopez+11,Rosen+14}). This ``X-ray suppression'' was suggested\footnote{It was alternatively suggested that the X-ray suppression in wind-blown bubbles is due to the bulk advection of hot gas energy out of the bubble, resulting in it's depressurization (e.g., \citealt{HarperClark&Murray09}) However, this explanation cannot explain the huge X-ray deficits in novae because the radiative cooling timescale of the hot gas is comparable or shorter than the outflow timescale.} to result from the hot shocked gas mixing with cooler gas, also present behind the radiative forward shock, after being swept up by the interstellar medium. As a result of efficient atomic cooling, lower temperature gas effectively instantaneously radiates any thermal energy it receives through this mixing process at UV/optical frequencies (e.g.,  \citealt{Pittard09,Kee+14,Rosen+14,Steinberg&Metzger18}). Using moving-mesh hydrodynamical simulations, \citet{Steinberg&Metzger18} showed how instabilities and mixing with cooler gas (gas swept up earlier which already cooled radiatively) can indeed suppress the X-ray luminosity by over an order of magnitude relative to the equivalent laminar case (see also \citealt{Kee+14}); however, given the enormous density contrast $\gtrsim 10^{3}$ between the hot and cold phases, numerically resolving the physical scales at which the mixing occurs remains extremely challenging (e.g., \citealt{Lancaster+24}), leaving it unclear whether the predicted levels of X-ray suppression are fully converged.

\begin{figure*}
    \centering
    \includegraphics[width=0.6\textwidth]{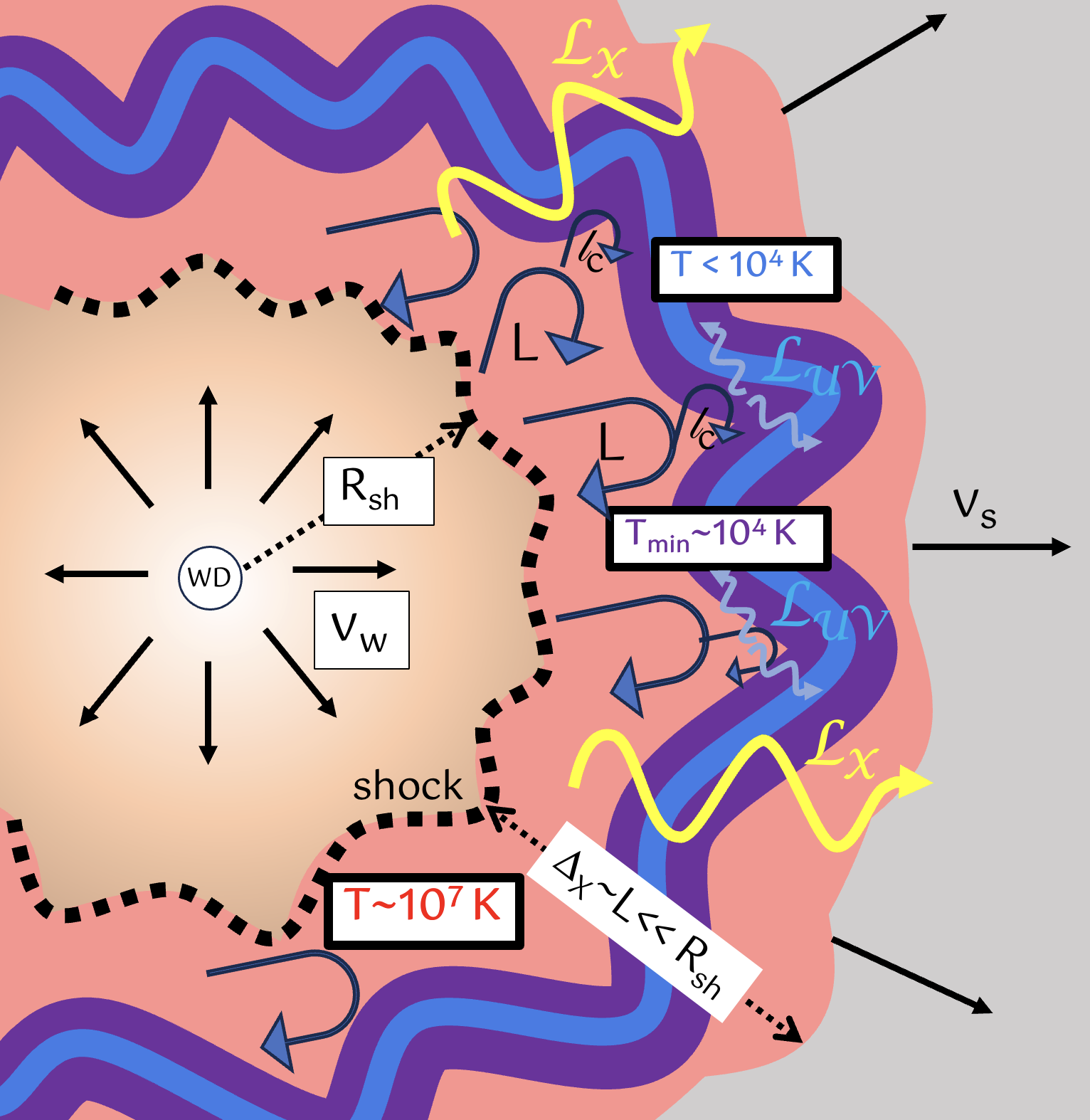}
    \caption{Illustration of how mixing behind the shocks in novae suppresses the X-ray luminosity. A fast outflow from the white dwarf of velocity $v_{\rm w} \gtrsim 1000$ km s$^{-1}$ collides with slower material released earlier of velocity $v_{\rm s} \ll v_{\rm w}$. By timescales at which X-ray emission has become visible of weeks to a month, this occurs at a characteristic radius $R_{\rm sh} \gtrsim 10^{14}$ cm. The bulk of the slow material resides in a cool, dense, and clumpy shell of temperature $T < 10^{4}$ K (blue region), which is the product of earlier slowly expanding nova ejecta that has cooled through radiation and expansion. The shock of velocity $v_{\rm sh} \approx v_{\rm w}$ heats the wind material to temperatures $T_{\rm X} \gtrsim 10^{7}$ K sufficient to radiate X-rays.  Inhomogeneities in the upstream medium and other instabilities render the shocked gas highly turbulent, with an outer turbulent length-scale $L$ comparable to the thickness of the X-ray emitting region, $\Delta_{\rm X}$. The hot shocked gas (red region) loses pressure rapidly by mixing with the cooler gas. This occurs through a turbulent fractal interface on a characteristic length-scale $\ell_{\rm c} \sim 10^{-4}L$ (not shown to scale) at which the eddy turnover timescale is comparable to the minimum radiative cooling timescale corresponding to mixed gas of intermediate temperature $T_{\rm c,min} \approx 10^{4}$ K (purple region). The cooler gas radiates the received energy effectively instantaneously UV light (which is at least partially absorbed and reprocessed by neutral gas into optical radiation), while the emitting volume and hence X-ray luminosity from the hot phase is greatly suppressed relative to the laminar case without mixing.}
    \label{fig:cartoon}
\end{figure*}

Given the challenges of numerical simulations, recent analytic work has focused on better understanding the physical processes responsible for energy exchange between hot and cool gas in turbulent mixing layers and the importance of the fractal nature of the hot-cold interface \citep{ElBadry+19,Fielding+20,Lancaster+21a,Lancaster+21b,Tan+21,Lancaster+24}.  In this paper, we present analytic estimates, motivated by these recent studies, to explore whether turbulent mixing between the hot and cold gas behind the radiative shocks in systems such as novae can explain their X-ray deficits.  In Sec.~\ref{sec:overview} we overview the properties of shocks in novae. In Sec.~\ref{sec:energy} we present a model for the X-ray emission from nova shocks, accounting for energy losses from the hot phase to turbulent mixing.  In Sec.~\ref{sec:discussion} we discuss some implications of our results. Sec.~\ref{sec:conclusions} summarizes our conclusions. The model is illustrated schematically in Fig.~\ref{fig:cartoon}, while Table \ref{tab:variables} summarizes the key variables to be introduced. 

\begin{deluxetable*}{ccc}
\tablecolumns{6}
\tablewidth{0pt}
 \tablecaption{Key Variables \label{tab:variables}}
 \tablehead{
 \colhead{Symbol}
 & \colhead{Definition}
 & \colhead{Estimated Range}
 }
 \startdata 
 $v_{\rm s}$ & velocity of slow initial outflow/cool shell & $200-600$ km s$^{-1}$ \\
 $\tau$ & optical-wavelength optical depth through slow outflow/cool shell & 1-10 \\
 $v_{\rm w}$ & velocity of fast white dwarf outflow & $1000-5000$ km s$^{-1}$ \\
 $\mathcal{L}_{\rm w}$ & kinetic power of fast outflow $\sim$ shock luminosity & $10^{37}-10^{38}$ erg s$^{-1}$  \\
$v_{\rm sh}$ & velocity of shock & $1000-5000$ km s$^{-1}$ \\
 $kT_{\rm X}$ & temperature of hot shocked gas & $1-10$ keV \\
 $R_{\rm sh}$ & radius of shock/shell at time of X-ray emission & $10^{14}-10^{15}$ cm  \\
 $P_{\rm sh}$ & pressure of shocked gas & $10^{-2}-10$ erg cm$^{-3}$ \\
$t_{\rm c,X}$ & radiative cooling time of shocked X-ray emitting gas & $10^{4}-10^{7}$ s \\
$\Delta_{\rm X}$ & radial thickness of X-ray emitting gas & $\sim$ min[$\Delta_{\rm rad}, R_{\rm sh}, L$]  \\
$\Delta_{\rm rad}$ & radial cooling length of X-ray emitting gas & $10^{-2}-1R_{\rm sh}$  \\
$L$ & outer length-scale of turbulent cascade & $10^{-5}-10^{-4}R_{\rm sh}$  \\
$A_{\rm sh}$ & (fractal) surface area separating hot and cool gas & $10^{3} \times (4\pi R_{\rm sh}^{2})$ \\
$\langle v_{\rm out} \rangle$ & advection speed of thermal energy from hot to cold gas & $10^{-2}v_{\rm w}$  \\
$f_{\rm t}$ & fraction of $\mathcal{L}_{\rm w}$ placed into post-shock turbulence & $10^{-3}-10^{-2}$  \\
$t_{\rm c,min}$ & minimum cooling time of turbulent mixed gas & $10-10^{3}$ s \\
$\ell_{\rm c}$ & turbulence length-scale dominating heat transport & $(10^{-5}-10^{-4})L$ \\ 
$\mathcal{L}_{\rm X}$ & X-ray luminosity of shocked gas & $10^{32}-10^{35}$ erg s$^{-1}$ \\
$f_{\rm X} \equiv \mathcal{L}_{\rm X}/\mathcal{L}_{\rm w}$ & X-ray radiative efficiency & $10^{-5}-10^{-3}$ \\
 \enddata
\end{deluxetable*}

\section{Overview of Shocks in Novae}
\label{sec:overview}

We consider a fast, approximately spherical wind from the white dwarf of mass-loss rate $\dot{M}_{\rm w} \sim 10^{-5}-10^{-4}M_{\odot}$ yr$^{-1}$, terminal velocity $v_{\rm w} \approx 1000-4000$ km s$^{-1}$ and kinetic luminosity $\mathcal{L}_{\rm w} = \dot{M}_{\rm w}v_{\rm w}^{2}/2$. We shall normalize $\mathcal{L}_{\rm w}$ to a value $\mathcal{L}_{\rm Edd} \sim 10^{38}$ erg s$^{-1}$ commensurate with the shock power one infers based on the observed gamma-ray luminosities $\mathcal{L}_{\gamma} \sim 10^{35}$ erg s$^{-1}$ (e.g., \citealt{Franckowiak+18}) around the time X-ray emission becomes visible in most novae, $t_{\rm X} \sim 10-30$ d (e.g., \citealt{Gordon+21}), for an assumed cosmic ray acceleration efficiency of $\sim 0.1\%$ (e.g., \citealt{Metzger+15}). Here, $\mathcal{L}_{\rm Edd} \approx 1.4\times 10^{38}$ erg s$^{-1}$ is the Eddington luminosity of a solar-mass white dwarf, which also represents the characteristic power of a wind accelerated by radiation pressure on the iron opacity bump \citep{Kato&Hachisu94,Shen&Quataert22}. 

The wind is assumed to collide with a comparatively slow and cool shell of velocity $v_{\rm s} \lesssim 500$ km s$^{-1} \ll v_{\rm w}$ and radius $R_{\rm s}$, which we also take to be spherical for simplicity.  This cool shell represents earlier mass-loss from the white dwarf, which has been swept up by the forward and reverse shocks and cooled through radiation or adiabatic expansion into a dense and geometrically-thin shell \citep{Metzger+14} of temperature $\sim 10^{3}-10^{4}$ K. This minimum temperature range is that expected from the balance between radiative cooling and heating from the nova light (e.g., \citealt{Cunningham+15}).
Assuming that most of the mass in the shell arises from the initial slow outflow launched at the beginning of the outburst (e.g., prior to optical maximum; \citealt{Aydi+20b}), by times of interest the shell has reached characteristic radii $R_{\rm sh} \sim v_{\rm s} t_{\rm X} \sim 10^{14}$ cm.

The wind-shell collision is mediated by a reverse shock of velocity $v_{\rm sh} \approx v_{\rm w} - v_{\rm s} \approx v_{\rm w}$, which heats the wind plasma to X-ray emitting temperatures, 
\be T_{\rm X} \simeq \frac{3}{16}\frac{\mu m_p}{k}v_{\rm sh}^{2} \simeq 1.4\times 10^{7}\,{\rm K}\,\left(\frac{v_{\rm w}}{10^{3}\,\rm km\,s^{-1}}\right)^{2},
\label{eq:TX}
\ee
and pressure \citep{Metzger+14}
\be
P_{\rm sh} \approx \frac{3}{4}\rho_{\rm w} v_{\rm sh}^{2}  \approx \frac{3}{16\pi}\frac{\mathcal{L}_{\rm w}}{v_{\rm w} R_{\rm sh}^{2}},
\label{eq:Psh}
\ee  
where $\rho_{\rm w} = \mu m_p n_{\rm w} = \dot{M}_{\rm w}/(4\pi R_{\rm sh}^{2}v_{\rm w})$ is the upstream density of the unshocked wind at radius $\approx R_{\rm sh}$ and $\mu \simeq 0.62$ is the mean molecular weight of ionized solar-composition gas. Here and in what follows we focus on emission from reverse shock, since its X-ray temperature and likely also its total power dominates that of the lower-velocity forward shock (e.g., \citealt{Metzger+14}).  However, most of the considerations to follow could just as well be applied to the forward shock.

At times of interest, the optical depth $\tau$ of the gas between the shock and an observer at infinity to optical wavelength radiation is sufficiently low that the photon diffusion time is shorter than the expansion time (i.e., $\tau < c/v_{\rm s} \sim 300$; e.g., \citealt{Arnett82}) and hence the gas cools as if it were effectively optically-thin. The radiative-cooling timescale of gas of pressure $P$ and number density $n = P/kT$ can be written,
\be
t_{\rm c} \equiv \frac{3}{2}\frac{P}{n^{2}\Lambda(T)} = \frac{3}{2}\frac{(kT)^{2}}{P\Lambda(T)},
\label{eq:tcool}
\ee
where $\Lambda(T)$ is the cooling function. Behind the shock where  $P \simeq P_{\rm sh}$ (Eq.~\eqref{eq:Psh}), the cooling time of the X-ray emitting gas can be estimated:
\begin{eqnarray}
t_{\rm c,X} &\equiv& \frac{3}{2}\frac{(kT_{\rm X})^{2}}{P_{\rm sh}\Lambda(T)} \approx 8\pi\frac{v_{\rm w}R_{\rm sh}^{2}(kT_{\rm X})^{2}}{\mathcal{L}_{\rm w}\Lambda(T_{\rm X})} \approx 8\times 10^{5}\,{\rm s}\,\frac{v_{\rm w,3}^{4}R_{14}^{2}}{\mathcal{L}_{\rm w,Edd}}\label{eq:tcoolsh}, \nonumber \\
\end{eqnarray}
where in the final line we have adopted short-hand notation $v_{\rm w,3} \equiv v_{\rm w}/(10^{3}$ km s$^{-1})$, $\mathcal{L}_{\rm w,Edd} \equiv \mathcal{L}_{\rm w}/\mathcal{L}_{\rm Edd}$, $R_{14} \equiv R_{\rm sh}/(10^{14}$ cm).
Here we have taken $\Lambda(T) \simeq 2\times 10^{-27} T^{1/2}$ cm$^{3}$ s$^{-1}$ for free-free emission, which dominates over atomic line-cooling for $T\gtrsim 10^{7}$ K ($v_{\rm sh} \gtrsim 10^{3}$ km s$^{-1}$). This timescale being comparable to the expansion time of the shocked gas $t_{\rm exp} \sim R_{\rm sh}/v_{\rm w} \sim 10^{6}$ s, has been used to argue that shock interaction in novae is radiative and hence approximately momentum-conserving \citep{Metzger+14}.  Because the shocked gas flows downstream away from the shock at a velocity $\approx v_{\rm sh}/4$, in a purely laminar picture (e.g. absent turbulent mixing described in the next section), it will cool radiatively over a characteristic radial distance,
\be \Delta_{\rm rad} \approx t_{\rm c,X}\left(\frac{v_{\rm sh}}{4}\right).
\label{eq:Deltac}
\ee
For high shock velocities, the reverse shock is not deeply radiative in the traditional sense that $\Delta_{\rm rad} \sim R_{\rm sh}$. However, we shall find below the interaction can still be {\it effectively} radiative due to turbulent mixing with cool gas, which has a much shorter radiative cooling time.

Another source of cooling for the hot gas, that could in principle suppress its X-ray luminosity, is Compton scattering off the nova's optical radiation field, as is known to be relevant in supernova ejecta undergoing circumstellar interaction (e.g., \citealt{Chevalier&Irwin12}).  Compton cooling of the X-ray emitting gas occurs on a timescale \citep{Rybicki&Lightman79}
\begin{eqnarray}
t_{\rm Y} \simeq \frac{3}{4}\frac{m_e c}{\sigma_{\rm T}U_{\rm rad}} = 3\pi \frac{R_{\rm sh}^{2}m_e c^{2}}{\mathcal{L}_{\rm w}\sigma_{\rm T}\tau}
\approx 8\times 10^{7}\,{\rm s}\,\frac{R_{14}^{2}}{\mathcal{L}_{\rm w,Edd}}\left(\frac{10}{\tau}\right),
\end{eqnarray}
where $m_e$ is the electron mass, $\sigma_{\rm T}$ is the Thomson cross section, $U_{\rm rad} \approx \mathcal{L}_{\rm opt}\tau/(4\pi R_{\rm sh}^{2}c)$ is an estimate of the energy density of the optical radiation field\footnote{This follows because the steady-state luminosity passing through a spherical shell of radius $r$ can be written as $L_{\rm opt} = 4\pi r^{2}U_{\rm rad}v_{\rm ph}$, where $v_{\rm ph} \approx c/\tau$ is the effective photon diffusion velocity when $\tau \gg 1$.} and in the second equality we have assumed most of the nova optical emission is reprocessed shock power, $\mathcal{L}_{\rm opt} \simeq \mathcal{L}_{\rm w}$ \citep{Li+17}.  The optical depth to the shock $\tau$ is typically modest $\tau \lesssim 10$ at epochs of interest (e.g., \citealt{Metzger+15}) and cannot exceed $\sim c/v_{\rm s} \sim 300$ for a collisionless shock capable of accelerating relativistic particles which emit the gamma-rays (e.g., \citealt{Levinson&Bromberg08}).  Insofar that $t_{\rm Y}$ is comparable or greater than the radiative cooling timescale (Eq.~\eqref{eq:tcoolsh}), we conclude that Compton losses alone cannot account for the huge X-ray deficits of nova shocks, which require $\mathcal{L}_{\rm X} \lesssim 10^{-4}\mathcal{L}_{\rm w}$.

\section{Suppressed X-ray Emission}
\label{sec:energy}

Within a 1D laminar picture, the hot shocked wind material and cool shell are separated by a smooth contact discontinuity across which pressure balance is established but no material flows.  However, this picture is too idealized, as the hot/cold boundary is instead a turbulent and fractal mess (e.g., \citealt{Steinberg&Metzger18}).  This turbulence can have many sources: shear created by temporal and/or angular variability in the white dwarf outflow, inhomogeneities in the upstream cold shell, and various hydrodynamic instabilities known to occur near or behind radiative shocks, particularly thin-shell \citep{Vishniac94} and thermal instabilities \citep{Field65,Chevalier&Imamura82}.  In this environment, the hot gas can lose thermal energy by turbulent mixing with cool gas faster than it radiatively cools on its own \citep{Steinberg&Metzger18,Fielding+20,Lancaster+24}, thereby suppressing the X-ray emission compared to a laminar radiative shock.  We aim to quantify this suppression here.

We consider a one-zone model for the shocked X-ray emitting gas, which we assume is localized in a shell of volume $V_{\rm X} = 4\pi R_{\rm sh}^{2}\Delta_{\rm X}$ and average radial thickness $\Delta_{\rm X} \ll R_{\rm sh}$. The total energy in the X-ray emitting gas, $E_{\rm X} = (3/2)P_{\rm sh}V_{\rm X}$,
evolves in time according to,
\be
\frac{dE_{\rm X}}{dt} = \mathcal{L}_{\rm w} - \mathcal{L}_{\rm X} - \dot{E}_{\rm mix}.
\label{eq:dEXdt}
\ee
The source term $\mathcal{L}_{\rm w}$ accounts for energy injected by the shock (the fast white dwarf outflow). The first sink term, 
\be
\mathcal{L}_{\rm X} \simeq \frac{E_{\rm X}}{t_{\rm c,X}},
\label{eq:LX}
\ee
accounts for energy radiated directly by the hot phase (as X-rays of characteristic energy $\sim kT_{\rm X}$; Eq.~\eqref{eq:TX}), where $t_{\rm c,X}$ is the cooling time of the hot gas (Eq.~\eqref{eq:tcoolsh}).

The second sink term in Eq.~\eqref{eq:dEXdt}, $\dot{E}_{\rm mix}$, accounts for turbulent mixing with the $\lesssim 10^{4}$K cooler gas; due to its much shorter cooling time, the mixed intermediate temperature gas radiates away the energy it receives from the hot gas effectively instantaneously, at optical/UV frequencies instead of X-rays.  
To estimate this term, we follow \citet{Fielding+20}, \citet{Lancaster+21a}, \citet{Lancaster+24}, and write the flux of energy through the cold-shell interface as:
\be
\dot{E}_{\rm mix} = \frac{5}{2}P_{\rm sh}A_{\rm sh}\langle v_{\rm out} \rangle, 
\label{eq:Edotmix}
\ee
where $\langle v_{\rm out} \rangle$ is an area-averaged effective speed at which hot plasma is transferred into the cold shell.  The surface area of the turbulent hot/cold interface can be written $A_{\rm sh} \approx 4\pi R_{\rm sh}^{2}/ \langle \hat{r}\cdot \hat{n} \rangle$, where $\langle \hat{r}\cdot \hat{n} \rangle \ll 1$ is the average value of the dot product between the radial and normal vectors over the fractal surface \citep{Lancaster+21b}.  As we shall discuss below, $A_{\rm sh} \gg 4\pi R_{\rm sh}^{2}$ as a result of the fractal nature of the surface \citep{Fielding+20}, substantially enhancing the mixing efficiency. Finally, by neglecting energy loss due to adiabatic expansion in Eq.~\eqref{eq:dEXdt}, we implicitly assume the shock interaction is radiative.

If turbulent mixing can be neglected ($\dot{E}_{\rm mix} = 0$) then in steady-state the kinetic luminosity of the shock is balanced by radiative cooling of the hot gas, i.e. $\mathcal{L}_{\rm X} = \mathcal{L}_{\rm w}$. In such a case, the shock's intrinsic (i.e., absorption-corrected) X-ray emission should roughly follow the shock power as measured by the gamma-ray light curve or the portion of the optical luminosity powered by reprocessed X-rays \citep{Li+17,Aydi+20}. However, this expectation is in strong tension with observations, for which $\mathcal{L}_{\rm X} \sim 10^{32}-10^{34}$ erg s$^{-1}$ at $t \sim 10-30$ d  (e.g., \citealt{Nelson+19,Sokolovsky+20,Gordon+21, Sokolovsky+22}), even while the optical and gamma-ray luminosities indicate ongoing powerful shocks $\mathcal{L}_{\rm w} \gtrsim 10^{37}-10^{38}$ erg s$^{-1}$.

On the other hand, if mixing with cool gas instead dominates over direct radiative cooling ($\dot{E}_{\rm mix} \gg \mathcal{L}_{\rm X}$) then
\be
\dot{E}_{\rm mix} \simeq \mathcal{L}_{\rm w}.
\label{eq:mixcondition}
\ee
As we will show below, in this limit the hot post-shock gas loses energy to mixing over a radial length-scale $\Delta_{\rm X} \ll \Delta_{\rm rad}$, thus greatly reducing the X-ray emitting volume compared to the laminar case (Eq.~\eqref{eq:Deltac}). Specifically, the X-ray luminosity in the rapid mixing limit can be written, (Eq.~\eqref{eq:LX})
\be
\mathcal{L}_{\rm X} = \frac{3}{2}P_{\rm sh}\frac{4\pi R_{\rm sh}^{2}\Delta_{\rm X}}{t_{\rm c,X}} = \frac{9}{32}\mathcal{L}_{\rm w}\frac{\Delta_{\rm X}}{\Delta_{\rm rad}},
\ee
where the second equality uses Eq.~\eqref{eq:Psh}.  A dimensionless ``X-ray efficiency'' can be defined,
\be
f_{\rm X} \equiv \frac{\mathcal{L}_{\rm X}}{\mathcal{L}_{\rm w}} \simeq \frac{9}{32}\frac{\Delta_{\rm X}}{\Delta_{\rm rad}}.
\label{eq:fX}
\ee
Efficient turbulent transport across the hot cold interface thus results in $\Delta_{\rm X} \ll \Delta_{\rm rad}$ and hence strongly suppressed X-rays $(f_{\rm X} \ll 1)$ relative to a laminar radiative shock ($\Delta_{\rm X} \sim \Delta_{\rm rad}$; $f_{\rm X} \simeq 1$).
Below we discuss how the minimum hot phase thickness $\Delta_{\rm X}$, and hence $f_{\rm X}$, may be set by the efficient mixing condition (Eq.~\eqref{eq:mixcondition}).

\subsection{Turbulent Diffusion Mixing}

We focus on energy transport due to hydrodynamic turbulent diffusion \citep{ElBadry+19,Fielding+20,Tan+21,Lancaster+21a}.  Although other physical effects, such as conduction and magnetic fields, can complicate this picture, we shall find they can generally be neglected to first order (see Sec.~\ref{sec:Bfields}). 

Instabilities driving turbulence behind the shock feed on the free energy of the flow.  We assume that a fraction $f_{\rm t}$ of the wind kinetic energy $\propto v_{\rm w}^{2}/2$ goes into turbulent motions of speed $\propto v_{\rm t}$ and kinetic energy $\propto v_{\rm t}^{2}/2$.  We take a canonical value $f_{\rm t} \sim 0.01$ motivated by the simulations of \citet{Steinberg&Metzger18}.\footnote{\citet{Steinberg&Metzger18} found that corrugation at the interface of a head-on radiative shock, imparted by thin-shell instabilities, results in $\sim 1\%$ of the incoming kinetic power being processed through oblique shocks capable of seeding vorticity and turbulent motions in the downstream.}  Turbulence is thus driven on some outer length-scale $L$ with a velocity amplitude:
\be
v_{\rm t}(L) = f_{\rm t}^{1/2} v_{\rm w}.
\ee
Below we show that when turbulent mixing is an important source of cooling behind nova shocks (i.e., to satisfy Eq.~\eqref{eq:mixcondition}), then $L$ is thin compared to the shock radius, i.e. $L \ll R_{\rm sh}$. Turbulent motions cascade to smaller scales $\ell < L$, following an assumed spectrum
\be
v_{\rm t}(\ell) = v_{\rm t}(L)\left(\frac{\ell}{L}\right)^{p} = v_{\rm t}(L)\left(\frac{\ell}{L}\right)^{1/3},
\label{eq:vt}
\ee
where we fiducially take $p = 1/3$ for subsonic, Kolmogorov turbulence.  The eddy turn-over time,
\be
t_{\rm eddy}(\ell) \equiv \frac{\ell}{v_{\rm t}(\ell)} \propto \ell^{1-p} = \ell^{2/3},
\label{eq:tedd}
\ee
becomes shorter for smaller scales $\ell \ll L.$ As described in \citet{Fielding+20}, \citet{Lancaster+21a}, the mixing that leads to cooling takes place on the scale $\ell_{\rm c}$ defined such that the mixing time $t_{\rm eddy}(\ell_{\rm c})$ equals the radiative cooling time $t_{\rm c}$ (Eq.~\eqref{eq:tcool}, but for the mixed temperature gas; see Eq.~\eqref{eq:tpk} below).  Using Eqs.~(\eqref{eq:vt},\eqref{eq:tedd}) we find
\be
\ell_{\rm c} \equiv L\left(\frac{t_{\rm c}}{t_{\rm eddy}(L)}\right)^{\frac{1}{1-p}} = L\left(\frac{t_{\rm c}}{t_{\rm eddy}(L)}\right)^{3/2},
\label{eq:ellc}
\ee
with an associated velocity scale,
\be 
v_{\rm t}(\ell_{\rm c}) = v_{\rm t}(L)\left(\frac{t_{\rm c}}{t_{\rm eddy}(L)}\right)^{\frac{p}{1-p}} = f_{\rm w}^{1/2}v_{\rm w}\left(\frac{t_{\rm c}}{t_{\rm eddy}(L)}\right)^{1/2} ,
\label{eq:vt_ellc}
\ee
where $t_{\rm eddy}(L) = L/v_{\rm t}(L)$ is the eddy overturn time at the outer scale. 

The specific cooling timescale $t_{\rm c}$ of relevance here refers not to that of the hot X-ray emitting gas introduced earlier (Eq.~\eqref{eq:tcoolsh}), but rather to that of the cooler gas at the cold-hot interface (e.g., \citealt{ElBadry+19,Lancaster+24}). Specifically, the key timescale for efficient mixing is the minimum cooling time $t_{\rm c} \propto T^{2}/\Lambda(T)$ (Eq.~\eqref{eq:tcool}) achieved at intermediate temperatures, as the hot gas is cooled through mixing with cold gas.  Assuming isobaric cooling at the post-shock pressure $P = P_{\rm sh}$, as expected given the subsonic nature of the post-shock turbulence, this minimum occurs for
\be
t_{\rm c,min} \equiv {\rm min}_{\rm T}(t_{\rm c}) = \frac{3}{2}\frac{(kT_{\rm c,min})^{2}}{P_{\rm sh} \Lambda(T_{\rm c,min})},
\label{eq:tpk}
\ee
at a temperature $T_{\rm c,min} \approx 1.6\times 10^{4}$ K for which $\Lambda(T_{\rm c,min}) \approx 2\times 10^{-22}$ erg cm$^{3}$ s$^{-1}$. Although line cooling dominates in this temperature range, the value of $t_{\rm c,min}$ is found to be insensitive to whether the ejecta possesses solar-composition ejecta or is enhanced in CNO elements from the white dwarf (see \citealt{Steinberg&Metzger18}, their Fig.~1).\footnote{In principle, photoionization by UV/X-ray radiation can alter the plasma cooling rates cooling relative to the LTE prediction (e.g., \citealt{Kim+23}); however, the ionization parameters of the UV/X-ray irradiated ejecta are typically modest at times and radii of interest \citep{Metzger+14}.}  

We can now estimate the quantities entering the rate of turbulent mixing (Eq.~\eqref{eq:Edotmix}). Heat is transported between the hot and cold gas at the velocity scale of turbulence capable of transporting gas faster than it can cool (e.g., \citealt{Fielding+20}), viz.~
\be
\langle v_{\rm out} \rangle \approx v_{\rm t}(\ell_{\rm c}).
\label{eq:vout}
\ee
The turbulent hot-cold interface is fractal in its geometry, strongly enhancing its surface area on small scales,
\begin{eqnarray}
A_{\rm sh}(\ell_{\rm c}) &=& 4\pi R_{\rm sh}^{2}\left(\frac{L}{\ell_{\rm c}}\right)^{d} \approx 4\pi R_{\rm sh}^{2}\left(\frac{t_{\rm c}}{t_{\rm eddy}(L)}\right)^{\frac{-d}{(1-p)}}  \nonumber \\
&=& 4\pi R_{\rm sh}^{2}\left(\frac{t_{\rm c}}{t_{\rm eddy}(L)}\right)^{-3/4},
\label{eq:Ash}
\end{eqnarray}
where $d$ is the so-called ``excess fractal dimension'' \citep{Fielding+20,Lancaster+21a}.  
The hydrodynamic simulations of \citet{Lancaster+24} are consistent with $d \approx 1/2$, which we adopt as the fiducial dimension hereafter.  
Because we shall find $L \ll R_{\rm sh}$, the geometry of the post-shock layer is effectively planar; unlike \citet{Lancaster+24}, we therefore neglect additional surface area enhancement between the spatial scales $L$ and $R_{\rm sh}$.

\subsection{Turbulent Cooling is Efficient}
\label{sec:mixing}

We can now address whether turbulent mixing behind the shock can efficiently cool the shocked gas, i.e. whether the condition $\dot{E}_{\rm mix} = \mathcal{L}_{\rm w}$ (Eq.~\eqref{eq:mixcondition}) can be satisfied for reasonable properties of the turbulence.

Using Eqs.~\eqref{eq:Psh}, \eqref{eq:vout}, \eqref{eq:Ash} from the previous section, the ratio of turbulence cooling to shock heating is given by,
\begin{eqnarray}
\frac{\dot{E}_{\rm mix}}{\mathcal{L}_{\rm w}} &=& \frac{5}{2}\frac{P_{\rm sh}A_{\rm sh}\langle v_{\rm out}\rangle}{\mathcal{L}_{\rm w}}  \approx \frac{15}{8}f_{\rm t}^{1/2}\left(\frac{t_{\rm c,min}}{t_{\rm eddy}(L)}\right)^{\frac{p-d}{1-p}} \nonumber \\
&=& \frac{15}{8}f_{\rm t}^{1/2}\left(\frac{L/v_{\rm t}(L)}{t_{\rm c,min}}\right)^{1/4} \nonumber \\
&\approx& 41 f_{\rm t,-2}^{3/8}\left(\frac{L}{R_{\rm sh}}\right)^{1/4}\frac{\mathcal{L}_{\rm w,Edd}^{1/4}}{R_{14}^{1/4}v_{\rm w,3}^{1/2}},
\label{eq:mixcondition2}
\end{eqnarray}
where $f_{\rm t,-2} \equiv f_{\rm t}/(0.01)$. Thus, even if the turbulence carries only a few percent of the available kinetic energy, with outer driving scale $L \gtrsim 10^{-6}R_{\rm sh}$ of only a tiny fraction of the shock radius, mixing with cool gas is sufficiently rapid to match the rate of hot gas creation at the shock, i.e. $\dot{E}_{\rm mix} \sim \mathcal{L}_{\rm w}$ can be satisfied.

\subsection{X-ray Emission}
\label{sec:Xrays}

The ease with which Eq.~\eqref{eq:mixcondition2} is satisfied for conditions characteristic of novae shows that turbulent mixing behind the shock is likely to be highly efficient at ``processing'' the hot gas created by the shock. This renders direct radiative cooling by the hot gas energetically irrelevant and implies a highly suppressed X-ray luminosity, $\mathcal{L}_{\rm X} \ll \mathcal{L}_{\rm w}$.  However, this then presents a challenge if one wants to calculate or even estimate the observed small value of $\mathcal{L}_{\rm X}$.  In particular, although mixing is efficient, it is not instantaneous.  In a 1D planar approximation for the post shock geometry, the short but finite time the hot gas spends radiating X-rays before mixing with cooler gas will set a characteristic radial thickness of the X-ray emitting layer, $\Delta_{\rm X}$, to which the X-ray luminosity is proportional (Eq.~\eqref{eq:LX}).  

If turbulent cooling balances shock heating, the condition $\dot{E}_{\rm mix} = \mathcal{L}_{\rm w}$ (Eq.~\eqref{eq:mixcondition2}) directly specifies a critical outer length-scale for the turbulence:
\be
L_{\rm min} = \left(\frac{8}{15}\right)^{\frac{1-p}{d-p}}\frac{v_{\rm w}t_{\rm c,min}}{f_{\rm t}^{\frac{1-d}{2(d-p)}}} = \left(\frac{8}{15}\right)^{4}f_{\rm t}^{-3/2}v_{\rm w}t_{\rm c,min}.
\label{eq:Lcrit}
\ee
We have labeled this length scale a minimum because if $L < L_{\rm min}$, then heating would exceed cooling ($\mathcal{L}_{\rm w} > \dot{E}_{\rm mix}$) and the size of the hot gas region separating the shock from the cool gas would expand.  This would potentially allow larger turbulent eddies to ``fit'' behind the shock, until $L \simeq L_{\rm min}$ and hence $\mathcal{L}_{\rm w} \approx \dot{E}_{\rm mix}$ becomes satisfied. On the other hand, if $L > L_{\rm min}$ and cooling exceeds heating ($\dot{E}_{\rm mix} > \mathcal{L}_{\rm w}$), then a similar feedback process could act to shrink the size of the post-shock region, and potentially reduce $L$. However, an alternative possibility when $L > L_{\rm min}$ is that even mixing which occurs on turbulent scales larger scales than $\ell_{\rm c}$ becomes sufficient to satisfy $\dot{E}_{\rm mix} \simeq \mathcal{L}_{\rm w}$. We return to this issue in Sec.~\ref{sec:nova_observations}.

The critical turbulence length $L_{\rm min}$ is small compared to the nominal radiative cooling length behind the shock $\Delta_{\rm rad}$ (Eq.~\eqref{eq:Deltac}),
\begin{eqnarray}
\frac{L_{\rm min}}{\Delta_{\rm rad}} &\simeq& 0.32 f_{\rm t}^{-3/2}\frac{t_{\rm c,min}}{t_{\rm c,X}} 
\simeq 0.32 f_{\rm t}^{-3/2}\left(\frac{T_{\rm min}}{T_{\rm X}}\right)^{2}\left(\frac{\Lambda(T_{\rm X})}{\Lambda(T_{\rm min})}\right) \nonumber \\
&\approx &  5\times 10^{-5}f_{\rm t,-2}^{-3/2}v_{\rm w,3}^{-3},
\label{eq:Lmin_ratio}
\end{eqnarray}
with a corresponding separation between the driving and cooling/mixing timescales given by:
\begin{eqnarray}
\frac{t_{\rm c,min}}{t_{\rm eddy}(L_{\rm min})} \approx \left(\frac{15}{8}\right)^{4}f_{\rm t}^{2} \approx 1.2\times 10^{-3} f_{\rm t,-2}^{2}.
\label{eq:tratio}
\end{eqnarray}
The small length scales over which turbulent transport occurs, $\ell_{\rm c}/L_{\rm min} = (t_{\rm c,min}/t_{\rm eddy}(L_{\rm min}))^{3/2} \sim 10f_{\rm t}^{3} \sim 10^{-5}$, highlights the challenges of resolving the mixing process with numerical simulations (e.g., \citealt{Kee+14,Steinberg&Metzger18,Lancaster+24}).

As the largest length-scale in the post-shock region, the driving scale $L$ also defines a minimum thickness for the hot X-ray emitting gas, i.e. $\Delta_{\rm X} \gtrsim L \gtrsim L_{\rm min}$. This is because hot gas created at the shock requires a volume corresponding to at least one outer eddy scale (i.e., the eddy must be able to `fit' behind the shock).  Because $L_{\rm min} \ll \Delta_{\rm rad}$, even a shock that would be only marginally radiative according to the usual laminar criterion (i.e., $\Delta_{\rm rad} \lesssim R_{\rm sh}$) can be deeply radiative as a result of turbulent cooling ($\Delta_{\rm X} \sim L_{\rm min} \ll R_{\rm sh}$) according to Eq.~\eqref{eq:Lmin_ratio}.

Using $\Delta_{\rm X} = L_{\rm min}$ in Eq.~\eqref{eq:fX}, we obtain a minimum on the X-ray radiative efficiency of the shock:
\be
f_{\rm X,min} \approx \frac{9}{32}\frac{L}{\Delta_{\rm rad}} \approx 1.4\times 10^{-5}f_{\rm t,-2}^{-3/2}v_{\rm w,3}^{-3},
\label{eq:fXfinal}
\ee
corresponding to an X-ray luminosity:
\begin{eqnarray}
\mathcal{L}_{\rm X,min} = f_{\rm X}\mathcal{L}_{\rm w} \approx 2\times 10^{33}\,{\rm erg\,s^{-1}}\,f_{\rm t,-2}^{-3/2}\mathcal{L}_{\rm w,Edd}v_{\rm w,3}^{-3}.
\label{eq:LX}
\end{eqnarray}
The predicted values $\mathcal{L}_{\rm X} \sim 10^{32}-10^{34}$ erg s$^{-1}$ for $\mathcal{L}_{\rm w} \sim \mathcal{L}_{\rm Edd}$, $f_{\rm t} \sim 1\%$, and $v_{\rm sh} \sim 500-5000$ km s$^{-1}$.  The fact that these values roughly match nova observations of $\mathcal{L}_{\rm X}$ may motivate the operation of the regulation process described above in which $L \approx L_{\rm min}$ and hence $\mathcal{L}_{\rm X} \approx \mathcal{L}_{\rm X,min}$(Sec.~\ref{sec:nova_observations}).
\section{Discussion}
\label{sec:discussion}

\subsection{Magnetic Fields and Conduction}
\label{sec:Bfields}

Turbulence can be suppressed by magnetic stresses if the velocity scale of the turbulence is less than the Alfv\'{e}n speed $v_{\rm A} = B/\sqrt{4\pi \rho}$ in the medium in question \citep{Chandrasekhar61}, where $B$ is the magnetic field strength.  This can have strong consequences if the mixing layer becomes magnetically dominated (e.g., \citealt{Lancaster+24}).

Magnetic fields must be present near the shocks in novae to give rise to the particle acceleration responsible for the gamma-ray emission (e.g., \citealt{Metzger+16,Diesing+23}). Insofar that the magnetic fields on the white dwarf surface are strongly diluted by flux freezing in the nova outflow, the required fields are likely amplified due to instabilities near the shock driven by the cosmic ray currents themselves (e.g., \citealt{Caprioli&Spitkovsky14a,Diesing&Caprioli19}).  For the fast shocks of relevance ($v_{\rm sh} \gtrsim 100$ km s$^{-1}$), the nonresonant instability dominates the magnetic field amplification \citep{Bell04}.  This results in a magnetic pressure close to the shock front (e.g., \citealt{Diesing+23}), 
\be
P_{\rm B} \simeq \frac{v_{\rm sh}}{2c}\frac{P_{\rm CR}}{\gamma_{\rm CR}-1},
\ee
where $P_{\rm CR}$ and $\gamma_{\rm CR} = 4/3$ are the cosmic ray pressure and adiabatic index, respectively.  Based on a comparison of the luminosities of the gamma-ray and (time-correlated, shock reprocessed) optical emission in classical novae, one infers that a fraction $\epsilon_{\rm p} \equiv P_{\rm CR}/P_{\rm sh} \sim 10^{-3}-10^{-2}$ of the kinetic power of the shock goes into accelerating cosmic ray ions \citep{Metzger+15,Li+17,Aydi+20}.  Taking $v_{\rm sh} \approx v_{\rm w}$, we can write,
\be
P_{\rm B} = \frac{3}{2}\epsilon_{\rm p}\frac{v_{\rm w}}{c}P_{\rm sh}.
\label{eq:PB}
\ee

Assuming Eq.~\eqref{eq:PB} holds, then on large scales $\sim L$, corresponding to the injection of the turbulence,
\be
\frac{v_{\rm t}^{2}(L)}{v_{\rm A}^{2}} = \frac{8}{3}f_{\rm t}\frac{P_{\rm sh}}{P_{\rm B}} = \frac{16}{9}\frac{f_{\rm t}}{\epsilon_{\rm p}}\frac{c}{v_{\rm w}} \gg 1,
\ee
and hence magnetic fields are dynamically irrelevant, where $v_{\rm A}^{2} = 2P_{\rm B}/\rho$. However, magnetic fields become dynamically more important on smaller scales further down the turbulent cascade. In particular, on the scales $v_{\rm t} \sim v_{\rm t}(\ell_{\rm c})$ (Eq.~\eqref{eq:vt_ellc}) which dominate turbulent energy diffusion, we find that:
\begin{eqnarray}
 \frac{v_{\rm t}^{2}(\ell_{\rm c})}{v_{\rm A}^{2}} 
\approx \frac{16}{9}\frac{f_{\rm t}}{\epsilon_{\rm p}}\frac{c}{v_{\rm w}}\frac{t_{\rm c,min}}{t_{\rm eddy}(L)} 
\approx 0.6 \frac{f_{\rm t,-2}^{3}}{v_{\rm w,3}}\left(\frac{10^{-3}}{\epsilon_{\rm p}}\right),
\end{eqnarray}
where in the final line we have used Eq.~\eqref{eq:tratio}. Here we have assumed $v_{\rm A} = constant$, when in reality flux-freezing in the compressing turbulent field leads to $B \propto \rho^{2/3}$ and hence $v_{\rm A} \propto \rho^{1/6}$ being larger in the cooled layer by a factor of a few given typical shock compression factors $\sim 10^{3}$. Our estimates suggest that the turbulence can be marginally magnetically-supported on the transport scale (i.e. $v_{\rm t}^{2} \gtrsim v_{\rm A}^{2}$) for particle-acceleration efficiencies $\epsilon_{\rm p} \sim 10^{-3}-10^{-2}$ needed to explain the gamma-ray emission \citep{Metzger+15,Li+17,Aydi+20}. This could imply that mixing would be limited to somewhat larger scales of the turbulent cascade and therefore be somewhat less efficient than we have assumed (e.g., \citealt{Lancaster+21b}).

On the other hand, the magnetic pressure on which these estimates have been made (Eq.~\eqref{eq:PB}) may be overestimated when the spectral properties of the gamma-ray emission are considered. To produce a maximum proton energy of $\sim 100$ GeV, consistent with that needed to explain the $\gamma$-ray emission of classical novae on a $\sim$ 1 day timescale \citep[e.g.,][]{Chomiuk+20}, one requires an acceleration time $\tau_{\rm acc} \simeq D/v_{\rm sh}^2 \simeq cr_{\rm L}/v_{\rm sh}^2 \simeq$ 1 day, where $D$ is the Bohm diffusion coefficient and $r_{\rm L}$ is the proton gyroradius. This requirement gives an approximate magnetic field $\simeq 1000 \mu$G in the vicinity of the shock, assuming $v_{\rm sh} = 1000 \ \rm{km \ s^{-1}}$. While this field certainly requires some sort of magnetic field amplification, it is either comparable to or lower than that calculated assuming Eq.~\eqref{eq:PB}, depending on our assumptions for $\epsilon_{\rm p}$ and $P_{\rm sh}$. This implies that the nonresonant instability might not saturate or might saturate at a lower level than that predicted by \cite{Bell04}, consistent with the results of kinetic simulations \citep{Zacharegkas+24}.

Thermal conduction can also transport energy from the hot to cool gas, and hence may compete with the turbulent transport of energy.  The balance between conduction and radiative cooling gives rise to a characteristic length-scale (e.g., \citealt{Field65,Lancaster+24})
\be
\lambda_{\rm F} = \sqrt{\frac{\kappa(T)T}{n^{2}\Lambda(T)}} = \sqrt{\frac{\kappa(T)T t_{\rm c}(T)}{P}},
\label{eq:lambdaF}
\ee
where $\kappa$ is the conductivity. Evaluating Eq.~\eqref{eq:lambdaF} for $T = T_{\rm c,min}$ and $P = P_{\rm sh}$, and using the estimate $\kappa \approx 1.6\times 10^{4}(T/1.6\times 10^{4}\,{\rm K})^{5/2}$ erg s$^{-1}$ cm$^{-1}$ K$^{-1}$ \citep{Parker53}, we find that
\begin{eqnarray}
\frac{\lambda_{\rm F}}{\ell_{\rm c}} &\approx& 
\left(\frac{\kappa(T_{\rm c,min})T_{\rm c,min}}{P_{\rm sh}L f_{\rm t}^{1/2}v_{\rm w}}\right)^{1/2}\left(\frac{t_{\rm c,min}}{t_{\rm eddy}(L)}\right)^{-1} \nonumber \\
&\simeq& 1.03 \left[\frac{\kappa(T_{\rm c,min})T_{\rm c,min} f_t^{3/2}\Lambda(T_{\rm c,min})}{(kT_{\rm c,min})^{2}v_{\rm w}^{2}}\right]^{1/2}\left(\frac{t_{\rm c,min}}{t_{\rm eddy}(L)}\right)^{-1} \nonumber \\
&\simeq& 0.64 f_{\rm t}^{-5/4}\left[\frac{\kappa(T_{\rm c,min}) \Lambda(T_{\rm c,min})}{k^{2}T_{\rm c,min}v_{\rm w}^{2}}\right]^{1/2}
\nonumber  \\
&\approx& 0.2 f_{\rm t,-2}^{-5/4}v_{\rm w,3}^{-1}.
\end{eqnarray}
The fact that $\lambda_{\rm F} \lesssim \ell_{\rm c}$ shows that thermal conduction becomes important on a smaller scale than the turbulent mixing occurs, and hence can be neglected in estimating the rate of heat transfer to first order.  The conductivity may also be suppressed relative to the assumed value by magnetic fields, provided the latter are draped across the interface between the hot and cold gas layers (e.g., \citealt{Orlando+08}). 

\subsection{Application to Nova X-ray Observations}
\label{sec:nova_observations}

\begin{figure}
    \centering
    \includegraphics[width=0.5\textwidth]{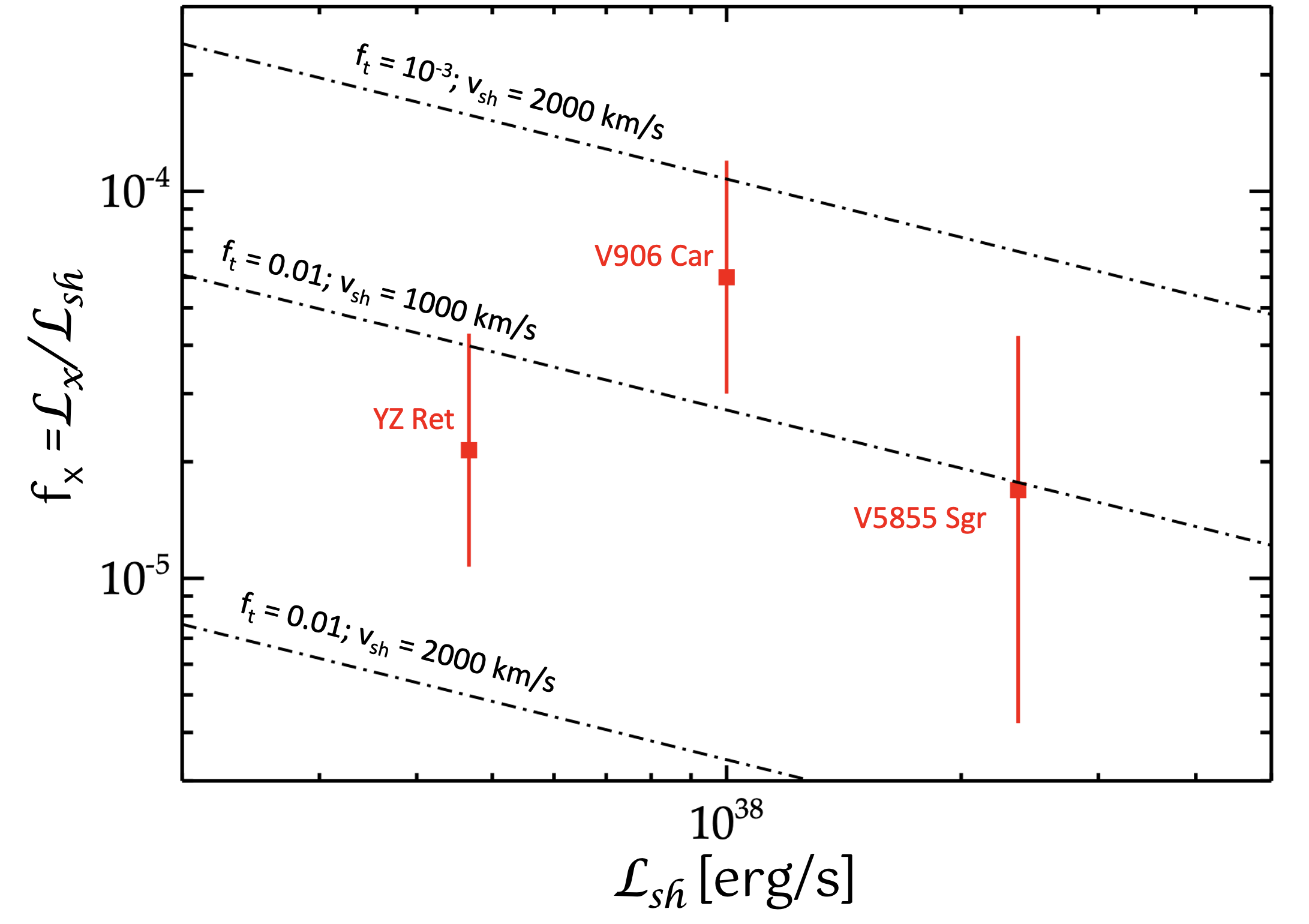}
    \caption{Predicted X-ray minimum suppression factor, $f_{\rm X,min} = \mathcal{L}_{\rm X,min}/\mathcal{L}_{\rm w}$ (Eq.~\eqref{eq:fXfinal}) as a function of shock luminosity $\mathcal{L}_{\rm w}$. Dot-dashed lines are shown for different parameter choices for the strength and outer length-scale of the turbulence and shock velocity as marked. Shown for comparison are data on $\mathcal{L}_{\rm X}/\mathcal{L}_{\rm w}$ from classical novae with contemporaneous (or nearly so) X-ray and gamma-ray detections, including YZ Ret \citep{Sokolovsky+22}, V906 Car \citep{Sokolovsky+20}, and V5855 Sgr \citep{Nelson+19}. We estimate the shock luminosity from the observed gamma-ray luminosity $\mathcal{L}_{\gamma} = \epsilon_{\gamma}\mathcal{L}_{\rm w}$, where a value $\epsilon_{\gamma} = 0.003$ was chosen motivated by correlated gamma-ray and optical light-curves \citep{Li+17,Aydi+20,Chomiuk+20}, assuming an order-unity fraction of the optical emission is reprocessed shock power \citep{Metzger+15}. Formal statistical error bars are shown on the data, but the systematic error bars exceed these by at least an order of magnitude.}
    \label{fig:fX}
\end{figure}

Figure \ref{fig:fX} shows our prediction for the minimum X-ray suppression factor $f_{\rm X}$ (Eq.~\eqref{eq:fX}) as a function of wind/shock luminosity $\mathcal{L}_{\rm w}$ for different values of the shock velocity and turbulent energy fractions as marked.  Shown for comparison are observed values of $f_{\rm X}$ for the small handful of novae with contemporaneous, or near contemporaneous, X-ray and gamma-ray detections \citep{Nelson+19,Sokolovsky+20,Sokolovsky+22}. For each event, we estimate the shock power from the observed gamma-ray luminosity $\mathcal{L}_{\gamma} = \epsilon_{\gamma}\mathcal{L}_{\rm w}$, where we take $\epsilon_{\gamma} = 0.003$ for the gamma-ray efficiency, motivated by correlated optical/gamma-ray emission in other novae assuming the gamma-ray tracking portion of the optical light curve is a proxy for the shock kinetic power \citep{Metzger+15,Li+17,Aydi+20}.  As a result of uncertainties in $\epsilon_{\gamma}$ and hence $\mathcal{L}_{\rm w}$, the true error on the ``observed'' $f_{\rm X}(\mathcal{L}_{\rm w})$ points clearly exceed the formal statistical errors on $\mathcal{L}_{\rm X}$ shown.

Despite these caveats, the data lie above our theoretical estimates of $f_{\rm X,min}$ (to within the order of magnitude uncertainties), provided that the shocks in novae place less than a percent of their available kinetic energy into turbulent motions (\citealt{Steinberg&Metzger18}).  
This rough quantitative agreement is notable given the simple nature of our estimates and several uncertain assumptions regarding the properties of the turbulence and the hot-cold interface which are challenging to calibrate from numerical simulations at high accuracy \citep{Lancaster+21b}.
For example, the predictions quoted here were made under fiducial assumptions about the turbulent cascade and fractal dimension ($p = 1/3$; $d = 1/2$), motivated by numerical simulations of turbulent mixing behind shocks (e.g., \citealt{Fielding+20,Lancaster+24}).  However, the dependence of the minimum X-ray luminosity (Eq.~\eqref{eq:LX}),
\be \mathcal{L}_{\rm X, min} \propto f_{\rm t}^{\frac{-(1-d)}{2(d-p)}},
\ee 
on the level of turbulence is sensitive to the small difference $d-p$. 

In Sec.~\ref{sec:Xrays} we were able only to estimate a minimum X-ray efficiency, by equating the thickness of the X-ray emitting layer behind the shock to the minimum outer scale of the turbulence, $L_{\rm min}$, necessary to cool gas at the rate of shock heating.  However, Fig.~\ref{fig:fX} shows that the observed X-ray luminosities are remarkably close to this predicted minimum value for fiducial parameters.  Insofar that $L_{\rm min}$ is much smaller than any other natural length scale behind the shock, this could appear to require fine-tuning.  However, it may instead hint at the existence of a regulated process, in which the outer turbulent scale is set by the thickness of the hot shock gas layer, which in turn is controlled by the balance between heating and cooling.  In particular, if $L > L_{\rm min}(L < L_{\rm min})$ then $\mathcal{L}_{\rm w} > \dot{E}_{\rm mix}$ ($\mathcal{L}_{\rm w} < \dot{E}_{\rm mix}$) and the thickness of the hot gas layer $\Delta_{\rm X}$, would expand (contract) until $\Delta_{\rm X} \approx L \approx L_{\rm min}$ is satisfied. Hydrodynamic simulations that explore the interplay between cooling and the processes driving turbulence behind radiative shocks are necessary to test this hypothesis.

\section{Conclusions}
\label{sec:conclusions}

The GeV gamma-ray emission from classical novae reveals the presence of powerful shocks of luminosity $\gtrsim 10^{37}-10^{38}$ erg s$^{-1}$ in these events \citep{Chomiuk+20}. However, the naive expectation that the shocked gas will emit most of its energy at X-ray energies corresponding to the immediate post-shock temperature is not borne out by observations \citep{Nelson+19,Sokolovsky+20,Chomiuk+20,Gordon+21,Sokolovsky+22}, with observed X-ray luminosities falling short of the shock power by roughly 4 orders of magnitude. A qualitative piece to the puzzle is clearly missing, which threatens to upend our basic understanding of the role that shocks play more generally in nova phenomenology \citep{Chomiuk+20}.

Here we have argued that a missing addition is the role played by turbulence in the post-shock gas, the effects of which are neglected in previous laminar shock interaction models (e.g., \citealt{Metzger+14}). Though a nearly ubiquitous feature of astrophysical fluid flows generally, in the context of novae turbulence is likely driven by temporal/spatial irregularities in the outflow, or thin-shell and thermal instabilities behind the reverse shock \citep{Steinberg&Metzger18}. The presence of a cool $T \lesssim 10^{3}-10^{4}$ K shell of gas in the nova ejecta is evidenced by the dust formation events observed to occur on timescales of weeks to a month similar to the observed X-rays (e.g., \citealt{Gordon+21}).

For typical assumptions about the driving turbulence, and the fractal dimension of the hot-cold interface motivated by hydrodynamic simulations  \citep{ElBadry+19,Fielding+20,Lancaster+21a,Lancaster+21b,Lancaster+24}, we find that turbulence can readily mix the hot gas with cooler gas much faster than the hot gas can radiate (Sec.~\ref{sec:mixing}). This loss of energy to cooler gas greatly suppresses direct emission from the hot gas, naturally explaining why nova shocks are so X-ray dim.

Although most of the shock's luminosity is transferred to the cooler gas, a small fraction is still emitted in X-rays, in proportion to the thickness of the hot gas layer behind the shock.  
We argue that a characteristic minimum thickness of this hot gas layer is the outer length scale of the turbulence $L$, and that $L$ must exceed a minimum value $L_{\rm min}$ to cool the hot gas through turbulent mixing at the rate it is heated by the shock.  A minimum X-ray radiative efficiency $f_{\rm X,min} \sim 10^{-5}-10^{-4}$ corresponding to $L \simeq L_{\rm min}$ was derived for typical shock and turbulence parameters (Eq.~\eqref{eq:fXfinal}). 

The fact that nova observations are consistent with $f_{\rm X} \sim f_{\rm X,min}$ for fiducial parameters (Fig.~\ref{fig:fX}) appears to suggest $L \sim L_{\rm min}$.  This coincidence is somewhat surprising because $L_{\rm min}$ is not a natural scale in the problem, being much smaller than either the shock radius or thickness of the cool shell.  This apparent fine-tuning $L \sim L_{\rm min}$ may instead hint at a presence of regulation process, in which the outer turbulence scale behind the shock matches that required to cool the gas through mixing at the same rate it is heated by the shock.  Whether and how such a feedback process, connecting the global thermodynamics with the turbulent driving, works is not yet clear, and probably can only be addressed with global hydrodynamic simulations.  

In this paper we have focused on emission from the reverse shock, since its X-ray temperature and total power likely dominate that of the lower velocity forward shock (e.g., \citealt{Metzger+14}).  However, most of the considerations regarding suppressed X-ray emission could equally well be applied to the forward shock, provided that the region separating the forward shock from the outer side of the cool shell is also turbulent.  The post-shock region may be considerably thinner when turbulent mixing dominates the cooling versus direct radiation of the hot gas. This would have implications for the synchrotron radio emission from novae \citep{Chomiuk+21}, which likely arises from the forward shock and has previously been modeled in a purely laminar shock picture (e.g., \citealt{Vlasov+16}).

Insofar that novae represent some of the closest examples of shock-powered transients in the universe, lessons learned from these events often have implications for more rarer and powerful extragalactic transients \citep{Fang+20}. For example, our model can in principle be applied to shock interaction in young supernovae, particularly the so-called Type IIn class which show strong circumstellar interaction (e.g., \citealt{Smith+07}). The earliest X-rays from Type IIn SNe can also originate from the radiative reverse shock \citep{Nymark+06}. However, compared to novae, the luminosities ($L_{\rm SN} \sim 10^{40}-10^{42}$ erg s$^{-1}$), velocities ($v_{\rm SN} \sim 5000$ km s$^{-1}$) and radii ($R_{\rm sh} \sim 10^{15}-10^{17}$ cm) of interacting supernova shocks are all substantially larger. Nevertheless, given the minimal requirements of our scenario, namely the presence of cool $T \lesssim 10^{4}$ K gas and a moderate level of turbulence, comparably large levels of X-ray suppression of several orders of magnitude, could also be expected in supernovae.

Whether observations of interacting supernovae support large X-ray suppression is unclear. The X-ray luminosities of young interacting core-collapse supernovae typically span a range $\mathcal{L}_{\rm X} \sim 10^{38}-10^{41}$ erg s$^{-1}$ (e.g., \citealt{Ross&Dwarkadas17,Dwarkadas25}) compatible with $f_{\rm X} \ll 1$ for powerful shocks $\mathcal{L}_{\rm w} \gtrsim 10^{40}-10^{43}$ erg s$^{-1}$; however, the intrinsic power of the reverse shock is challenging to determine independently in most events. Nevertheless, our study motivates further explorations of the impact of turbulent mixing on the emission from interacting supernovae and other shock-powered transients.

\vspace{12pt}
\noindent {\bf Acknowledgments}

We thank the anonymous reviewer for comments which helped improve the manuscript. BDM thanks Eliot Quataert for helpful suggestions motivating this work and Taya Govreen-Segal for conversations. We also thank Vikram Dwarkadas for comments on the manuscript. BDM acknowledges support from NASA AAG (grant number 80NSSC22K0807), the Fermi Guest Investigator Program (grant number 80NSSC24K0408) and the Simons Foundation (grant number 727700). L.L. gratefully acknowledges the support of the Simons Foundation under grant 965367.



\bibliographystyle{mn2e}
\bibliography{ms}

\end{document}